# Spontaneous Symmetry Breaking and Dynamic Phase Transition in Monolayer Silicene

Lan Chen, Hui Li, Baojie Feng, Zijing Ding, Jinglan Qiu, Peng Cheng, Kehui Wu* and Sheng Meng*

*Institute of Physics, Chinese Academy of Sciences, Beijing 100190, China*

*Corresponding Authors: khwu@iphy.ac.cn (K.H. Wu) & smeng@iphy.ac.cn (S. Meng).

The (√3×√3)R30° honeycomb of silicene monolayer on Ag(111) was found to undergo a phase transition to two types of mirror-symmetric boundary-separated rhombic phases at temperatures below 40 K by scanning tunneling microscopy. The first-principles calculations reveal that weak interactions between silicene and Ag(111) drive the spontaneous ultra buckling in the monolayer silicene, forming two energy-degenerate and mirror-symmetric (√3×√3)R30° rhombic phases, in which the linear band dispersion near Dirac point (DP) and a significant gap opening (150 meV) at DP were induced. The low transition barrier between these two phases enables them interchangeable through dynamic flip-flop motion, resulting in the (√3×√3)R30° honeycomb structure observed at high temperature.

Silicene, a sheet of Si atoms arranged in a honeycomb lattice analogous to graphene [1], does not exist in nature but has been theoretically predicted [2-4], and successfully prepared on metal [5, 6] and semiconductor surfaces recently [7]. Experiments reveal similarities between silicene and graphene: honeycomb structure, linear dispersion of the electron band as well as high Fermi velocity ($10^6$ m/s) [3, 4]. The stronger spin-orbit coupling (SOC) in Si than in C is also intriguing since it may induce possible quantum spin-Hall effect (QSHE) [4, 8] and quantum anomalous Hall effect (QAHE) [9]. Due to the large Si-Si bond length and partial $sp^3$ hybridization, free-standing silicene maintains a non-planar, so-called low-buckled (LB) geometry with reduced symmetry as compared with 1×1 graphene [3, 4]. Experimentally, silicene grown on Ag(111) surface exhibits a variety of different structural phases such as 4×4 [6, 10-13], √13×√13 [6, 11-13], √7×√7 [6, 13], 2√3×2√3 [12, 13] (with respect to Ag(111) surface lattice) and √3×√3 [5, 6] (with respect to silicene 1×1). The existence of Dirac Fermions was confirmed in the (√3×√3)R30° (simplified as √3) superstructure through the observation of quasiparticle interference (QPI) patterns in scanning tunneling microscopy (STM) dI/dV maps [5]. However, extensive first-principles calculations so far, with or without the Ag(111) substrate, have not be able to reproduce the √3 superstructure [4, 14]. As a result, the atomic arrangements in the honeycomb √3 phase structure remain illusive. It is a challenge that one has to overcome in order to further explore novel physics of silicene such as QSHE and to realize potential spintronics applications.

In this Letter, the origin of the peculiar √3 phase of silicene on Ag(111) is resolved through an experimental discovery of a dramatic low temperature phase transition in silicene. Below 40 K we observed a spontaneous symmetry breaking of silicene sheet on Ag(111) surface, by forming two mirror-symmetric rhombic √3 superstructures. In calculations, we show that it is the dispersive interaction between Si atom and Ag substrate that drives the formation of the two mirror-symmetric√3 superstructures. The ultra buckling of Si atoms in √3 structure induce a large gap (150 meV) at the Dirac point, which may help to realize detectable QSHE and other device applications. Our results not only elucidate the origin of √3×√3 superstructure and the nature of honeycomb to rhombic phase transitions in epitaxial silicene, but also provide profound implications that weak dispersive interactions might tip the balance among various competing epitaxial structures and result in rich interface phenomena.

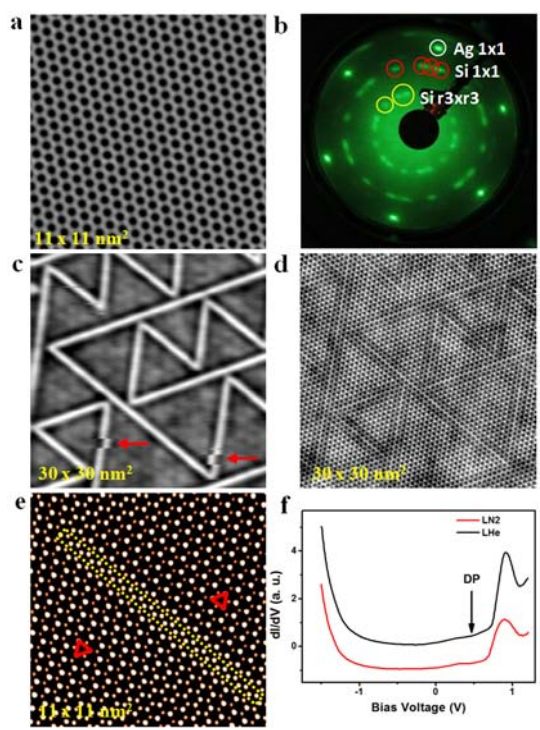

Fig. 1 (a) The high resolution STM image of monolayer silicene taken at tip bias 1.0V at 77 K. (b) LEED pattern taken at 51.4 eV on √3 ×√3R30° silicene. The white, red and yellow circles represent the Ag(111) 1×1 spot, silicene 1×1 spots and silicene √3 ×√3 ones, respectively. (c) and (d) The STM images of the same area on √3 ×√3R30° silicene taken at tip bias -2.0 V and 0.1 V, respectively, at 5 K. (e) The filtered high-resolution STM image with high contrast taken at 0.1V. The yellow rectangle marks the position of domain boundary. The red triangles indicate the brighter protrusions in one unit cell in different domains. (f) dI/dV curves taken at 77 K (red curve) and 5 K (black curve). The position of Dirac point (DP) is labeled.

The experimental conditions and sample preparation were identical to that in Ref. 5. The monolayer silicene film on Ag(111) exhibits a honeycomb structure with a period of 0.64±0.01 nm (Fig. 1(a)), corresponding to (√3×√3)R30° superstructure with respect to the silicene 1×1 lattice, for sample temperature at 77 K or up to room temperature (RT). The (√3×√3)R30° structure can be confirmed by the low energy electron diffraction (LEED) pattern shown in Fig.1(b). We found that, there are four predominant orientations of silicene with reference to the [1-10] direction of Ag(111): 0°, ±10°, and 30°. More intriguingly, when the sample is cooled to liquid Helium temperature (5 K), a dramatic structural phase transition occurs, which is characterized by the appearance of atomic chains forming interconnected triangles (Fig.1(c)). A close inspection reveals that these are boundaries separating two symmetric domains, as shown in Fig.1(d). At 77 K, the two neighboring protrusions in each honeycomb unit cell are equally bright. While upon the phase transition, one of them becomes much brighter than the other showing an apparently rhombic √3 superstructures (Fig.1(e)). As there are two possible configurations, the surface is phase separated into triangular domains with either one of the two symmetric configurations, separated by narrow domain boundaries where the neighbor protrusions are identically bright , as indicated by the yellow rectangle in Fig.1(e)). Temperature-dependent experiments show that the phase transition takes place at about 40 K.

We performed scanning tunneling spectroscopy (STS) measurements on silicene to investigate electronic structures before and after the phase transition. Typical dI/dV curves obtained at 77 K and 5 K (Fig. 1(f)) reveal similar electronic structures, both with a small dip located at about 0.5 V attributed to the position of Dirac point (DP) of silicene, and a pronounced peak at 0.9 V,. The triangular domain boundaries can serve as quasiparticle scattering centers to result in pronounced standing wave patterns in STS maps (Fig. 2(a)). Typical dI/dV maps at different bias voltages are shown in Fig. 2(b), 2(c) and 2(d), exhibiting strong QPI patterns due to intravalley scattering in single Dirac cone (see Fig. 2(e)). In order to deduce the quasiparticle energy-momentum dispersion relation, we drew $E(\kappa)$ curve in Fig. 2(f), where $\kappa$ is the radius of constant-energy circle at K point with $2\kappa=|q_1|$, and $q_1$ is intravalley scattering wave vector. The values of $q_1$ are determined by measuring the wave length of QPI patterns in dI/dV maps. We found $\kappa$ varied linearly with energy, with Fermi velocity $V_F = (0.97 \pm 0.02) \times 10^6$ m/s. The $\kappa=0$ energy intercept gives the Dirac energy, $E_F-E_D = 0.50 \pm 0.02$ eV, in consistency with the position of DP in dI/dV spectra (Fig. 1(f)) very well. The electronic structure of silicene is similar to each other for the low temperature and high temperature phases, except a slightly smaller Fermi velocity for the phase at 5 K than that at 77 K (($1.2 \pm 0.1) \times 10^6$ m/s, see Ref.5).

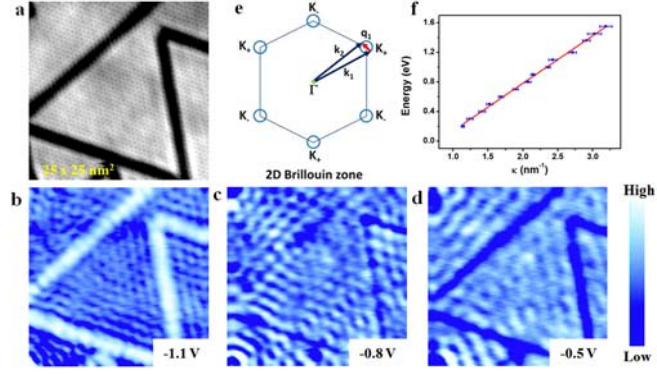

Fig. 2 (a) The STM image of silicene with domain boundaries taken at tip bias -1.0 V. (b), (c) and (d) dI/dV maps of the same area as (a) taken at tip bias -1.1 V, -0.8 V and -0.5 V, respectively. (e) Schematic of 2D Brillouin zone (grey lines), constant energy contours (blue rings) at K points, and intravalley scattering vectors $q_1$ (red arrow). (f) Engery dispersion as a function of κ (defined in text) for silicene determined from wave length in dI/dV maps. The red line shows a linear fit to the data.

The observed √3 superstructure, as well as temperature-induced phase transition is unique for silicene, which is absent in other 2D materials such as graphene [1] and boron nitride (BN) [15]. Previously, extensive first-principles calculations have all pointed out that free standing silicene monolayer adopts a general low-bulked (LB) geometry [3, 4], in which three alternative atoms in a hexagonal ring are buckled upward, leading to 1×1 structure in STM images. However, the √3 superstructure has never been reproduced so far by first principles calculations with or without the Ag substrate [10, 11, 14],. In a previous work, we had proposed a phenomenological, compressed √3 model, which features double side buckling [5]. However, this model does not form spontaneously on Ag(111), and it failed to explain the low-temperature phase transition either. It should be also noted that the √3 superstructure is not from the lattice commensuration between the silicene and Ag(111), since there are four different orientations of silicene with respect to Ag(111), as shown in Fig. 1(b).

It is well-known that both conventional Local Density Approximation (LDA) and General Gradient Approximation (GGA) in the density functional theory (DFT) framework cannot accurately describe the non-covalently bonded silicene-substrate interaction, which may play a critical role in the formation of the √3 superstructure. Fortunately, dispersion corrected DFT methods have been developed in recent years, such as density functionals that could account for nonlocal van der Waals (vdW) interactions, namely, the vdW density functionals (vdW-DF) [16]. The vdW-DF approaches have been successfully applied in a variety of adsorption systems [17, 18] with a better reproduction of geometry and interaction energy of layered materials and molecules adsorbed on metals. Therefore, in our present calculations we include the dispersive interaction [19] and perform structural search for Si monolayer adsorbed on Ag(111) [20].

To match the experimental observations, we adopted a variety of prototype supercells to model epitaxial silicene, each having a distinct orientation with respect to the underlying Ag(111) lattice: (i) A (3×3) silicene supercell on a (4×4) Ag(111) slab with five Ag atomic layers, of which the primary cell vector of (1×1) silicene makes a relative orientation angle $\theta = 0°$ with respect to the [1-10] direction of Ag(111) surface lattice; (ii) A (√21×√21)R10.9° silicene cell on a (6×6) surface cell of Ag(111) with the orientation angle $\theta = 10.9°$; (iii) A (3√3×3√3)R30° silicene cell on Ag(111) (7×7) supercell with the orientation angle $\theta = 30°$.

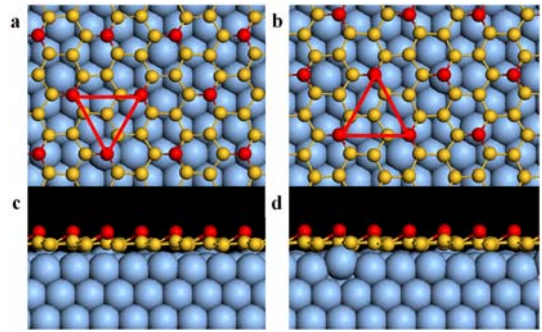

Fig. 3 The top views (a, b) and side views (c, d) of two energy-degenerated √3 reconstructed structures of silicene sheet on Ag(111) surface with an orientation angle of $\theta = 30°$, which are obtained from DFT optimization in the case of both lattices of √3 silicene and silver substrate having identical direction. Color code: Blue, yellow, and red spheres denote Ag atoms, Si atoms in lower layer, and Si atoms in higher layer, respectively. The red triangles in (a) and (b) denote the units of √3 silicene structures.

In all the three cases, silicene sheets spontaneously relaxed into √3 superstructure with vdW correction involved in the calculations [20]. Since the relaxed structural models are similar among the three cases, the silicene layer with the orientation angle $\theta$ = 30° is chosen as an example to illuminate the calculated √3 superstructure. As shown in Fig. 3(a) and 3(b), two mirror-symmetric (√3×√3) reconstruction structures have been found with the same adsorption sites of Si atoms on Ag(111). These two √3 superstructures have identical geometry if ignoring the substrate, and share the same central substrate atom for each hexagon unit (red triangles in Fig. 3). In each √3×√3 unit cell, only one Si atom is buckled upward (red spheres in Fig. 3), whereas the other five Si atoms have almost same lower height (yellow spheres in Fig. 3), resulting in rhombic (√3×√3)R30° superstructure. The rhombic √3 silicene has much larger buckling height (about 1.2 Å) than that in free standing (1×1) silicene (0.4 Å) [3], as well as smaller Si(upper)-Si(lower) bond angles (102-107°), indicating a stronger $sp^3$ hybridization involved. The two types of rhombic √3 superstructures accord with the two mirror-symmetric phases observed in low temperature experiments (Fig. 1(e)) very well. We note that in the rhombic √3 structure, these buckled Si atoms locate on the hollow, bright or even atop sites of silver substrate, implying the interaction between silicene and Ag(111) are not position-specific. This proves the crucial role of weak vdW interactions, and is consistent with the appearance of several orientations of silicene in experiments. The √3 phases forms by balancing between maximizing Si-Ag chemical bonds and maximizing the interlayer vdW interactions. At the same time, the intralayer strain within silicene layer is released by maintaining an average Si-Si bond length of 2.31 Å.

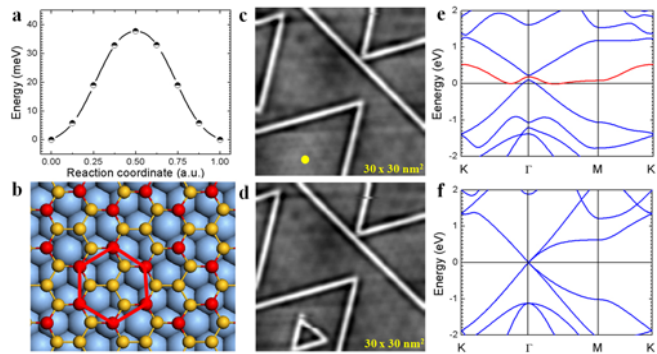

Fig. 4 (a) The interpolated potential energy curve for structural transition between the two the mirror-symmetric √3 geometries on Ag(111). (b) The intermediate structure between the two rhombic √3 structures of silicene shown in Fig. 3(a) and (b). (c) and (d) Experimetal STM images of same area taken at -2.0V before and after a 3.0 V pulse (10 ms) applied. The yellow dot marks the position of pulse applied. (e) and (f) The band structures of √3 structure of silicene layer and 1×1 structure of free standing silicene, respectively.

Based on our DFT calculation, the two mirror symmetric rhombic √3 superstructures have the same binding energy. Using simple interpolation-scanned potential energy surface, the phase transition between the two structures is found to be quite easy, with a roughly estimated barrier of ≤ 38 meV per Si atom, as shown in Fig. 4(a). Such a low barrier allows the phase transition to be thermo-activated at temperatures of dozens of Kelvin. That is to say, the six Si atoms around the center of hexagon will be flip-floped so quickly that the STM scanning cannot follow such changes. As a consequence, a *honeycomb* √3 superstructure would be observed at higher temperatures (shown in Fig. 4(b)). The similar temperature-induced phase transition have been known for Si(100)-1×2 [21] and Si(111)-Ag(√3×√3) surface [22, 23]. In STM experiments at 5K, we found that the domain boundary may be lateral shifted suddenly during scanning (shown by the arrows in Fig. 1(c)), which means the two kinds of rhombic √3 structures can be transited into each other through the activation by electronic field between the tip and substrate. Additionally, we tried to control this structure transition through applying a bias pulse on the center of a domain, and then found a small new trianglular domain with an opposite mirror symmetry formed at the pulsing position (Fig. 4(c) and Fig. 4(d)).

The reduced crystalline symmetry and the stronger buckling of Si atoms are believed to influence the electronic structures of silicene. We calculated the band structure of the √3 superstructure (without Ag), as shown in Fig. 4(e). For comparison, the band structure of unreconstructed silicene 1×1 phase was also displayed in Fig. 4(f). Similar to the unreconstructed phase, the $\pi_1$ and $\pi_1^*$ bands of the reconstructed phases, which are contributed from $p_z$ orbitals of Si atoms in lower layer, maintains the linear energy-momentum dispersion at Γ point (K and K' points in Brillioun zone of (1×1) phase are folded onto Γ point in the Brillioun zone of (√3×√3)R30° superstructure) and charge carriers behave as Dirac Fermions. Moreover, a significant gap at the DP (~0.15 eV) is opened for the √3 phase. This gap opening should be the result of the high buckling reconstruction in √3 phase, which largely distorts the delocalized π molecular orbital parallel to the silicene surface. However, STS reflects the total local density of states (LDOS) of the sample, and the LDOS of underlying Ag(111) substrate contribute to STS strongly and may cover the gap at DP. So such a big gap was not observed in dI/dV curves in STM experiments. It is interesting to note that there is a flat band (red line in Fig. 4(e)) going through the center of the gap at DP in the √3 phase. This band is mainly contributed by the $p_z$ atomic orbitals of three Si atoms next around the high buckled atom, and will be filled with electrons transferred from Ag(111) substrate when silicene is adsorbed onto Ag(111). As a consequence, the σ-bond-like interaction between low-layer Si atoms and Ag substrate will be formed, which is an important factor for the stabilization of the √3 silicene monolayer adsorbed on Ag(111). Indeed, the √3 superstructure is spontaneously relaxed back into (1×1) phases without Ag substrate in the calculations.

The dynamic √3 phase of silicene on Ag(111) is energetically stable, and contains linear dispersion in the electronic bands. Therefore it must share the same nontrivial topological properties as the free standing, low-buckled silicene. The first principles calculation shows the gap induced by ultra buckling of Si is larger than 100 meV, corresponding to a temperature higher than room temperature, making silicene an ideal system for future device applications. The stronger buckling of √3 superstructure than that of free-standing 1×1 phase may result in stronger SOC, which will help to realize the QSHE [4] and QAHE [9]. Moreover, understanding the structure of silicene on metal surfaces will build a base for investigating other important physical phenomena such as superconductivity [24, 25]. In experiment, the synthesis of large area silicene on insulating substrate with a high dielectric constant will be the next challenge.

**Acknowledgements:** This work was financially supported by the MOST of China (Project No. 2012CB921703 and 2012CB921403), and the NSF of China (Project No. 11174344, No. 11074289, No. 91121003, No. 11222431 and No. 11074287). L. Chen and H. Li acknowledge the

start funding from IOP, CAS.